\documentclass[aps,showpacs,showkeys]{revtex4}
\usepackage{amsmath,amssymb,bm,amsthm}
\usepackage{graphicx,subfigure,enumerate}
\usepackage{dcolumn}% Align table columns on decimal point
\usepackage{longtable}
\usepackage[dvipdfm,colorlinks=true,citecolor=blue,pdfstartview=FitH]{hyperref}

\def\bea{\begin{equation}}
\def\eea{\end{equation}}
\newcommand{\rt}{Regge trajectory}
\newcommand{\rts}{Regge trajectories}
\newcommand{\tj}{trajectory}
\newcommand{\tjs}{trajectories}

\begin{document}
\title{Concavity of the meson Regge trajectories}

\author{Jiao-Kai Chen}
\email{chenjk@sxnu.edu.cn, chenjkphy@outlook.com}
\affiliation{School of Physics and Information Engineering, Shanxi Normal University, Linfen 041004, China}
\date{\today}

\begin{abstract}
It is illustrated by the fitted Regge trajectories for a large majority of mesons that the radial and orbital Regge trajectories for mesons prefer being concave in the $(n_r,M^2)$ and $(l,M^2)$ planes, respectively. The concavity of the meson Regge trajectories is model-independent.
The concavity is expected to act as a criterion to choose a newly observed meson or to assign a particle to the unwell-established state.
The concavity of the meson Regge trajectories can assist in model construction and in the choice of the appropriate model to describe mesons. The appropriate models should yield the spectra which can produce the concave Regge trajectories according to the concavity of the meson Regge trajectories. If the large majority of the meson Regge trajectories are concave while a few meson Regge trajectories are convex which neither have been confirmed nor have been completely excluded at present, many existing models should be corrected or even be reconstructed, which will lead to the further understanding of the meson dynamics.
\end{abstract}

%12.40.Nn Regge theory, duality, absorptive/optical models 每
%12.39.Pn Potential models
%11.10.St Bound and unstable states;Bethe-Salpeter equations 每
%12.40.Yx Hadron mass models and calculations
%14.40.-n Mesons (for leptonic decays of mesons, see 13.20.-v; for hadronic decays of mesons, see 13.25.-k)
%Use showkeys class option if keyword
                              %display desired
\maketitle
%%%%%%%%%%%%%%%%%%%%%%%%%%%%%%%%%%%%%%%%%%%%%%%%%%%%%%%%%%%%%%%%%%%%%%%%%%

\section{Introduction}\label{sec:int}
The Regge theory \cite{Regge:1959mz,Collins:1971ff} is concerned with the particle spectrum, the forces between particles, and the high energy behavior of scattering amplitudes; in fact with almost all aspects of strong interactions. As one of the most distinctive features of the Regge theory, the {\rt} is one of the effective approaches for studying hadron spectra \cite{Regge:1959mz,Collins:1971ff,Chew:1961ev,Hey:1982aj,Inopin:2001ub,Chen:2016spr,
Chen:2016qju,Inopin:1999nf,Chew:1962eu}. The {\rts} are essentially non-linear complex functions. Their complexity is required by several reasons: (1) the finite widths of resonances require it; (2) the analytic S-matrix theory and the Regge pole model in particular require threshold singularities in the scattering amplitude, consequently the {\tjs} produce the imaginary parts; (3) while the imaginary part of the {\tj} may rise indefinitely, its real part is limited \cite{Bugrij:1973ph}. Complex meson and baryon {\tjs} were fitted to both resonances's masses and widths \cite{Fiore:2000fp,Fiore:2004xb,Silva-Castro:2018hup,Fiore:2015lnz,Fernandez-Ramirez:2015fbq}.
The spectrum can be described by the real parts of the {\tjs} while the resonance widths are related to the imaginary parts. We restrict ourselves to the real parts of the complex {\rts} in this work, which are hereinafter referred to as the {\rts}. The famous Chew-Frautschi plots of the {\rts} provide a useful way of hadron classification. Linearity is a convenient guide in constructing the Chew-Frautschi plots and the linear {\rts} have played an important role in studying hadron spectra. On the other hand, various studies have shown that the {\rts} can be nonlinear \cite{Tang:2000tb,Brisudova:1999ut}.
The {\rts} can be divided into two groups depending on whether they are linear or not. One group is the linear {\rts} \cite{Chew:1961ev,Chew:1962eu,Nambu:1974zg,Nambu:1978bd,Ademollo:1969nx,Karch:2006pv,Baker:2002km,Polchinski:2001tt,
Brodsky:2006uq,Forkel:2007cm,Filipponi:1997vf,Klempt:2012fy,Afonin:2007aa,Ebert:2017els,Brodsky:2018vyy,Masjuan:2017fzu,Baldicchi:1998gt,Anisovich:2000kxa,Masjuan:2012gc}. Another group of the {\rts} are nonlinear.

For heavy quarkonia especially for bottomonia, the {\rts} deviate significantly from the linearity \cite{Chen:2018hnx,Sharov:2013tga,Wei:2010zza,Brisudova:1998wq}.
The nonlinearity of the {\rts} for light mesons, for heavy-light mesons and for baryons are also investigated \cite{Inopin:1999nf,Tang:2000tb,Chen:2018nnr,Sharov:2013tga,Sonnenschein:2014jwa}.
The nonlinear {\rts} take different forms. In Refs. \cite{frau:1968rg,Martinis:1969ch,maryab69,vasa214,vasa2442,vasa453,grib93}, the authors show that the effective {\tj} for large momentum transfer $(-t)$ goes like $\sqrt{-t}$. In Ref. \cite{Lyubimov:1977km}, the author presents the nucleon {\tj} obtained from the backward scattering of pions by nucleons, $\alpha_N(t)=-0.4+0.9t+0.25t^2/2$. In Ref. \cite{Brandt:1997gi}, UA8 proposes the effective Pomeron {\rt} by analyzing the inclusive differential cross sections for the single-diffractive reactions $p_i+\bar{p}{\to}p_f+X$ and $p+\bar{p}_i{\to}X+\bar{p}_f$ at $\sqrt{s}=630$ GeV, $\alpha_P(t)=1.10+0.25t+(0.079\pm0.012)t^2$, where $t$ is the squared-momentum-transfer. See Refs. \cite{Inopin:2001ub,Inopin:1999nf} for more {\rts}.

In Ref. \cite{Inopin:1999nf}, the authors scrutinize the hadron {\rts} in the framework of the string and potential models, and they observe the appreciable curvature of the orbital {\rts} for mesons and baryons. The fitted {\rts} which have the inflection points are not observed in the references. The nonlinear {\rts} are expected to be concave or convex. In Refs. \cite{Dey:1994za,hendry:1980tk}, the flattening of the resonance spectrum of hadrons is observed, i.e., the hadron {\rts} are concave in the $(J,M^2)$ plane. In Ref. \cite{PandoZayas:2003yb}, the authors notice the curvature of the {\rts}, $\alpha''(t)>0$. The {\rts} are concave for the heavy mesons \cite{Chen:2018hnx,Wei:2010zza} and for many light mesons and baryons  \cite{Sharov:2013tga,Sonnenschein:2014bia}.
In Refs. \cite{Sharov:2013tga,Ebert:2011jc,Sonnenschein:2014jwa,Afonin:2013hla,Branz:2010ub,Wei:2010zza,
Gershtein:2006ng,Afonin:2014nya}, the fitted curves show evidently the concavity although the authors do not point out explicitly the curvature of the meson {\rts}. In Refs. \cite{Brisudova:1999ut,Inopin:1999nf,
Brisudova:1998wq}, the authors notice the curvature of the {\rts} for mesons but only a few {\rts} are fitted.
We conclude that the generality of the concavity of the meson {\rts} has not been pointed out clearly and has not been demonstrated either.
In the present work, we illustrate by the fitted {\rts} for a large majority of mesons that the meson {\rts}, both the radial {\rts} and the orbital {\rts}, prefer being concave and this property is model-independent.
The effects of the concavity of the {\rts} on hadron classification and model construction are also discussed.

This paper is organized as follows. In Sec. \ref{sec:rts}, the concavity of the meson {\rts} is illustrated. In Sec. \ref{sec:gprop}, the general properties of the {\rts} and the effects of the concavity of the meson {\rts} on model construction and hadron classification are discussed. In Sec. \ref{sec:comp}, the complexity of the {\rts} is presented. We conclude in Sec. \ref{sec:con}.

\section{Concave meson {\rts}}\label{sec:rts}
In this section, we show that the radial and orbital {\rts} for the large majority of mesons can be well described by the concave formulas. Some {\rts} for the light unflavored mesons are not definite due to scarce data or the unwell-established states. We propose that the radial and orbital meson {\rts} prefer being concave in the $(n_r,M^2)$ and $(l,M^2)$ planes, respectively.

\subsection{Concavity of the {\rts} for heavy quarkonia}

In Ref. \cite{Chen:2018hnx}, we propose one new form of the {\rts} from the quadratic form of the spinless Salpeter equation \cite{Chen:2018hnx,baldicchi:reg,chen:2016gvs,chen:2017qsse,DiSalvo:1994mf,Baldicchi:2007ic,Baldicchi:2000cf,Brambilla:1995bm} by employing the Bohr-Sommerfeld quantization approach \cite{brau:04bs,tomonaga},
\begin{align}\label{rtint}
M^2=&\beta_l(l+b_l)^{2/3}+c_l, \nonumber\\
M^2=&\beta_{n_r}(n_r+b_{n_r})^{2/3}+c_{n_r},
\end{align}
where $M$ is the meson mass, $l$ is the orbital angular momentum, $n_r$ is the radial quantum number. $\beta_l$, $\beta_{n_r}$, $b_l$, $b_{n_r}$, $c_l$ and $c_{n_r}$ are parameters. For heavy quarkonia, $\beta_l$ and $\beta_{n_r}$ are universal parameters.

Applying the concave formulas (\ref{rtint}) to fit the spectra of bottomonia and charmonia, the fitted {\rts} are in good agreement with the experimental data and the theoretical predictions \cite{Chen:2018hnx}. The {\rts} will be concave if the spectra can be described by the concave formulas, therefore, the {\rts} for heavy quarkonia are concave. The concavity of these {\rts} is independent of the models from which the employed formulas come.

\subsection{Concavity of the {\rts} for the mesons consisting of different quarks}
In Ref. \cite{Chen:2018nnr}, we show that the {\rts} (\ref{rtint}) are also appropriate for the mesons composed of unequally massive quarks. We employ Eq. (\ref{rtint}) to fit the spectra of the strange mesons, the heavy-light mesons (the $D$, $D_s$, $B$ and $B_s$ mesons) and the bottom-charmed mesons. The fitted {\rts} are consistent with the experimental data and the theoretical values. Thus, the {\rts} for these mesons are concave because the data can be described by the concave formulas. The concavity of the {\rts} is model-independent.

\subsection{Concavity of the {\rts} for the light unflavored mesons}\label{sub:concave}
For the light unflavored mesons, there are eight radial {\rts} and ten orbital {\rts} can be described by Eq. (\ref{rtint}) which are concave. There are eight radial {\rts} and eight orbital {\rts} seem convex. No convex {\rts} for the light unflavored mesons have been confirmed until now. More theoretical analyses and more experimental data are needed to confirm or exclude the possible convexity of the {\rts}. See Appendix \ref{app:light} for more discussions.

The parameters $\beta_l$ and $\beta_{n_r}$ are universal parameters for heavy quarkonia and the mesons constituting of unequally massive quarks. But for the light unflavored mesons, the universal description of these two parameters does not hold again, see Tables \ref{tab:coefr} and \ref{tab:coefo}.

\subsection{Convex {\rts}}\label{sub:convex}

\begin{figure*}[!hpbt]
\centering
\subfigure[]{\label{fig:subfigure:3a}
\includegraphics[scale=0.6]{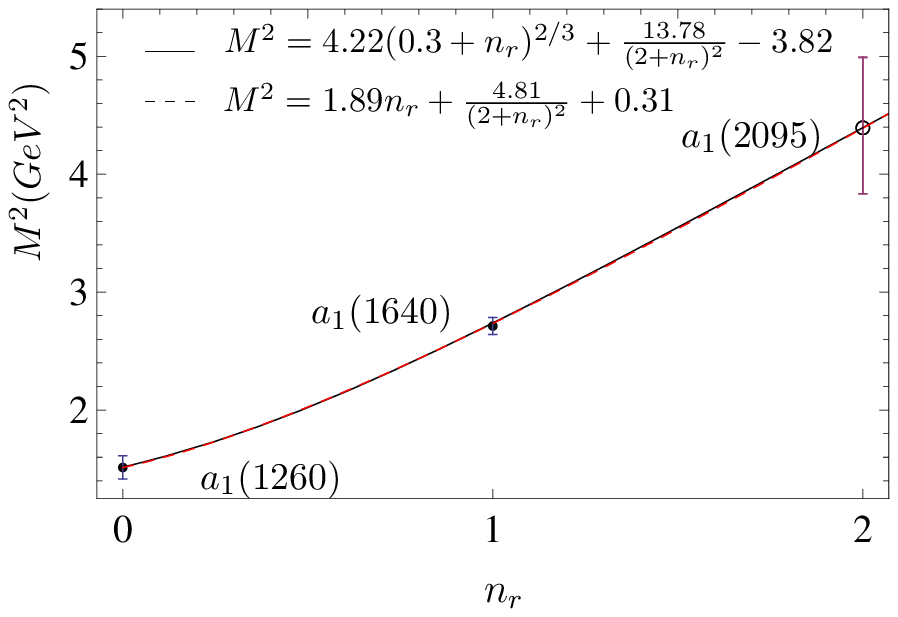}}
\subfigure[]{\label{fig:subfigure:3b}
\includegraphics[scale=0.6]{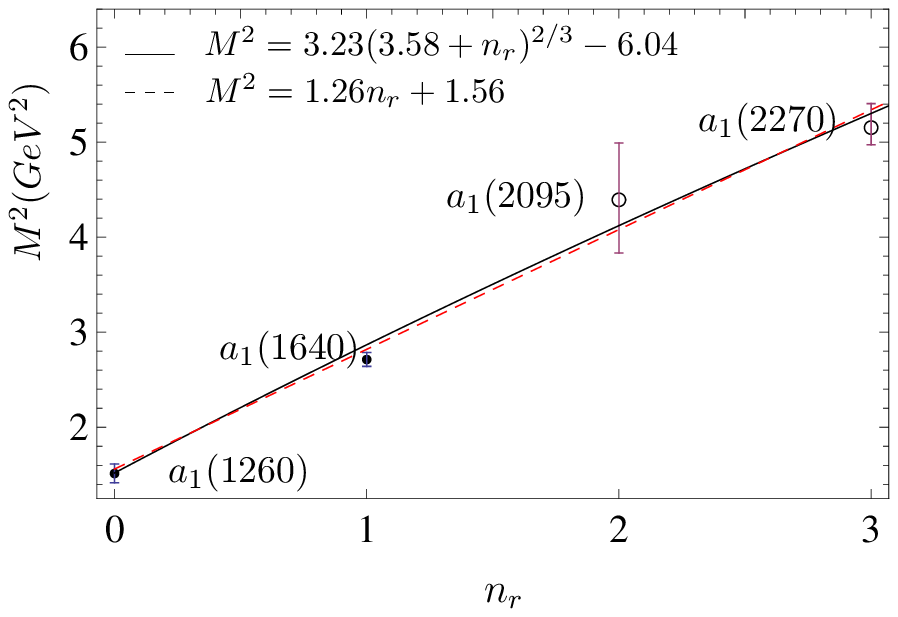}}
\subfigure[]{\label{fig:subfigure:3c}
\includegraphics[scale=0.6]{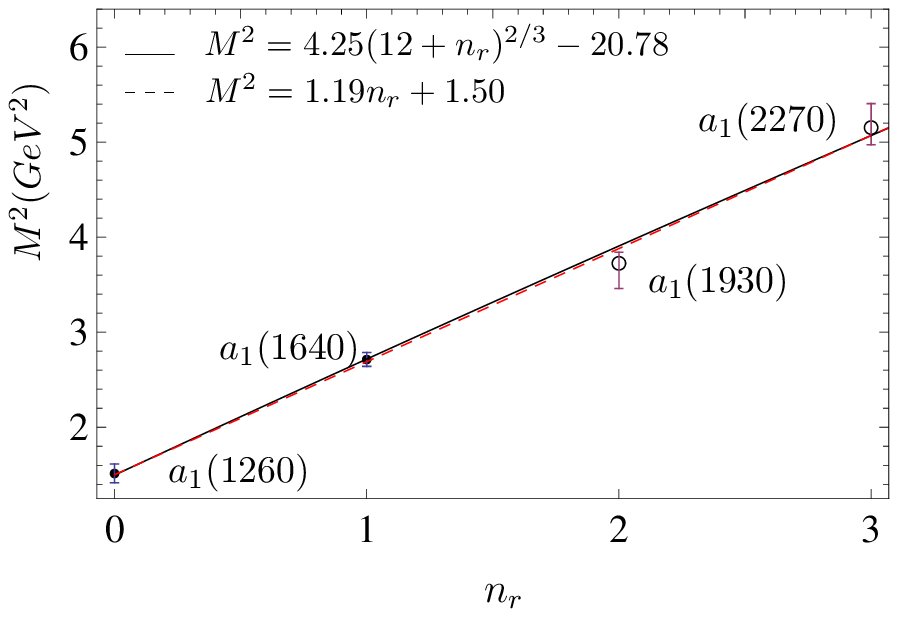}}
\caption{(Color online) The radial {\rts} for the $a_1$ mesons. The data are from Ref. \cite{Tanabashi2018pdg}. The well-established states are given by solid dots and the candidates are given by circles.}\label{fig:cv}
\end{figure*}

Different from the concave {\rts} mentioned above, the {\rts} for some light unflavored mesons look like convex. In this subsection, we show by example that the convex {\rts} are problematic.

Takes as an example, the radial {\rt} for $a_1(1260)$, $a_1(1640)$ and $a_1(2095)$ is discussed. $a_1(1260)$, $a_1(1640)$ are the well-established states, assigned to be the $1^3P_1$ and $2^3P_1$ states, respectively. In Refs. \cite{Masjuan:2012gc,Ebert:2009ub,Sonnenschein:2014jwa}, $a_1(2095)$ is assigned to be the $3^3P_1$ state. The {\rt} for these three states are convex and cannot be described by the concave {\rts} (\ref{rtint}). Using the interpolated formula in Eq. (\ref{nrtf}), the fitted results are consistent with the experimental data, see Fig. \ref{fig:subfigure:3a}. However, the fitted parameter $\gamma_2$ is positive which should be negative. It is not acceptable physically because the first term in Eq. (\ref{nrtf}) comes from the color Coulomb potential and should be negative. The same problem exists for the {\rt} formula \cite{Sergeenko:1993sn,Sergeenko:1994ck,Ebert:2011jc}
\bea\label{nonlreg}
M^2=\tau_1(J+2n_r)+\frac{\tau_2}{(n_r+J+1)^2}+\tau_0,
\eea
where $\tau_1$ should be positive while $\tau_2$ should be negative. The fitted {\rt} for the $a_1$ mesons by using Eq. (\ref{nonlreg}) is also unacceptable because $\tau_2>0$, see Fig. \ref{fig:subfigure:3a}.

If the {\rt} for the $a_1$ mesons is really convex, Eqs. (\ref{nrtf}) and (\ref{nonlreg}) can be the temporary formulas to fit the data. If the correct {\rt} for the $a_1$ mesons is concave, the improper {\rt} maybe results from the insufficient experimental data or the inappropriate assignment of states. As an example of the insufficient data case, the experimental data can be described by the concave formula as the candidate state $a_1(2270)$ is considered, see Fig. \ref{fig:subfigure:3b}. In the latter case, the {\rt} becomes concave if not $a_1(2095)$ but $a_1(1930)$ is assigned as $3^3P_1$ state \cite{Chen:2015iqa}, see Fig. \ref{fig:subfigure:3c}. Some other {\rts} for the light unflavored mesons are in this case. Up to present, the convex {\rts} for mesons have not been definitely confirmed and have not been completely excluded either.

\subsection{Discussions}
The dynamics of the light mesons is more complicated than that of heavy quarkonium because the light mesons are rather sensitive to detailed properties of the confinement mechanism. Therefore, the fitted {\rts} for heavy mesons will be better than the {\rts} for the light mesons. All fitted {\rts} for heavy quarkonia are concave \cite{Chen:2018hnx}. All the bottom-charmed mesons, the heavy-light mesons (the $D/D_s/B/B_s$ mesons) and the strange mesons can be described by the concave {\rts} [Eq. (\ref{rtint})] \cite{Chen:2018nnr}. Most of the {\rts} for the light unflavored mesons are concave. Therefore, the large majority of the {\rts} for mesons are concave \cite{Inopin:1999nf,Brisudova:1999ut,Sharov:2013tga,Sonnenschein:2014jwa,Afonin:2013hla,Branz:2010ub,Chen:2018nnr}. On the other hand, no {\rts} having the inflection points are observed and only a few convex {\rts} are given in references \cite{Klempt:2009pi,deTeramond:2005su,Nandan:2016uce,Ranjan:2011ge}. Up to date, no convex {\rts} for mesons are definitely confirmed. Combining the discussions in this section and in Appendix \ref{app:light}, we can conclude that the radial and orbital meson {\rts} prefer being concave in the $(n_r,M^2)$ and $(l,M^2)$ planes, respectively. The concavity of the meson {\rts} is model-independent although the used formulas are obtained from different models.

The curvature of the meson {\rts} can be used to classify hadrons. For mesons, the concavity can act as a criterion to choose a newly observed meson or to assign a particle to the unwell-established state.
If the great majority of the meson {\rts} are concave while a few meson {\rts} are convex which have not been confirmed and have not been completely excluded either at present, mesons can be classified according to different curvature.

\section{Effects of the concavity of the meson {\rts} on models}\label{sec:gprop}
In this section, the general properties of the {\rts} are presented. The effects of the concavity of the mesons {\rts} on model construction are discussed.

\subsection{General properties of the {\rts}}
Based on the meson spectra obtained experimentally \cite{Tanabashi2018pdg} and predicted theoretically \cite{Eichten:1974af,Eichten:1979ms,Godfrey:1985xj,Ebert:2009ub,Ebert:2009ua,Ebert:2011jc,Liu:2016efm,
Baldicchi:2007zn,Pang:2017dlw}, it is evident that the orbital {\rts} $M(l)$ and the radial {\rts} $M(n_r)$ should be the strictly increasing functions,
\bea\label{regdv}
\frac{dM^2(l)}{dl}>0,\;\; \frac{dM^2(n_r)}{dn_r}>0,
\eea
where $M$ is the meson mass, $l$ is the orbital angular momentum and $n_r$ is the radial quantum number. Let $\kappa$ be the curvature of the {\rts}. If the {\rts} are linear, $\kappa=0$,
\bea
\frac{dM^2(l)}{dl}={\rm const.},\;\;\frac{dM^2(n_r)}{dn_r}={\rm const.}.
\eea
If the {\rts} are convex upwards in the $(l,M^2)$ and $(n_r,M^2)$ planes, $\kappa>0$,
\begin{align}\label{rtconvex}
\frac{d^2M^2(l)}{dl^2}>0,\;\;\frac{d^2M^2(n_r)}{dn_r^2}>0.
\end{align}
If the {\rts} are concave functions, $\kappa<0$,
\begin{align}\label{rtconcave}
\frac{d^2M^2(l)}{dl^2}<0,\;\;\frac{d^2M^2(n_r)}{dn_r^2}<0.
\end{align}
Considering Eqs. (\ref{regdv}) and (\ref{rtconcave}), we can obtain
\bea\label{condinf}
\lim_{l\to\infty}\frac{d^2M^2(l)}{dl^2}=0,\;\;\lim_{n_r\to\infty}\frac{d^2M^2(n_r)}{dn_r^2}=0
\eea
if the variables $l$ and $n_r$ can be extended to infinity. If the {\rts} are concave, they should satisfy Eqs. (\ref{regdv}), (\ref{rtconcave}) and/or (\ref{condinf}). If the {\rts} are convex, they should meet Eqs. (\ref{regdv}) and (\ref{rtconvex}).

Eq. (\ref{condinf}) does not hold if $l$ and $n_r$ have the upper bounds, for example, the {\rt} $t=-0.139 l^2+2.55 l+115.98$ listed in \cite{Brisudova:1999ut}, in which $l$ should be smaller than or equal to the maximum value $l_{max}$.
If the {\rts} have the inflection points on the domain, they will change from being concave to being convex or vice versa. This kind of {\rts} are not favored because the changing from being concave to being convex suggests the possible changing of dynamics. In fact, no meson {\rts} having the inflection points have been observed. The {\rts} $M^2=\tau_1(J+2n_r)+{\tau_2}{(n_r+J+1)^{-2}}+\tau_0$ [Eq. (\ref{nonlreg})] \cite{Sergeenko:1993sn,Sergeenko:1994ck,Ebert:2011jc} and  $M^2={\gamma_1}{(l+n_r+1)^{-2}}+\beta_1(l+b_1)^{2/3}+c_1$ [Eq. (\ref{nrtf})] can be concave, convex or have the inflections if the physical constraints on the parameters are not considered. As the physical constraints on the parameters are considered, they should be concave.

\subsection{Dynamic equations}\label{subsec:dyn}
The potential models are the basic tools of the phenomenological approach to model the features of QCD relevant to hadron with the aim to produce concrete results \cite{Eichten:1978tg,Eichten:2007qx,Brambilla:2004wf,Lucha:1991vn,Kwong:1987mj,Godfrey:1985xj,Fulcher:1991dm,
Ebert:2009ub,Ebert:2009ua,Ebert:2011jc,Liu:2016efm,Bali:2000gf,Gupta:1993pd}. Different dynamic equations and different potentials will lead to different {\rts} \cite{Chen:2018hnx,Chen:2018nnr,Kahana:1993yd,regKG,
shar1185,martin:86r,olsson:94regd,Goebel:1989sd,Lucha:1991vn,DiSalvo:1994mf,Badalian:2016ttl,FabreDeLaRipelle:1988zr,
baldicchi:reg,Sergeenko:1993sn,Sergeenko:1994ck}. The concavity of the meson {\rts} is of significance for the potential models, because it can assist in the choice of the appropriate dynamic equation and the corresponding potential.

The concavity of the {\rts} gives the constraints on the confinement potential if the dynamic equation is ascertained. For the Schr\"{o}dinger equation, the confinement potential should be $r^a$ with $0<a<2/3$ \cite{brau:04bs,FabreDeLaRipelle:1988zr}. For the Dirac equation \cite{olsson:94regd}, the spinless Salpeter equation \cite{brau:04bs,Lucha:1991vn,Goebel:1989sd,martin:86r} and the Klein-Gordon equation \cite{Sergeenko:1993sn,Sergeenko:1994ck,Goebel:1989sd,regKG,shar1185}, the confinement potential should be $r^a$ with $0<a<1$. For the quadratic form of the spinless Salpeter-type equation \cite{Chen:2018hnx} or the eigenvalue equation of the square mass operator \cite{Chen:2018hnx,DiSalvo:1994mf}, the confinement potential should take the form $r^a$ with $0<a<2$.

If the confinement potential is $r^{a}$ with $0<a<2/3$, all the dynamic equations mentioned above can produce the concave {\rts}. In this case, we need more information to choose the appropriate dynamic equation. For example, the Martin potential $-8.064+6.898r^{0.1}$ \cite{Martin:1980jx,Eichten:1994gt}, the Song-Lin potential $-Ar^{-1/2}+kr^{1/2}+V_0$ \cite{Song:1986ix} and the Motyka-Zalewiski potential $-Ar^{-1}+kr^{1/2}+V_0$ \cite{Motyka:1995ng,Motyka:1997di} are in this case. $A$ is the strong interaction constant, $k$ is the confinement constant.

\subsection{Linear confinement potential}\label{subsec:linear}
The popular confinement is a linearly rising potential which has been validated by lattice QCD calculations \cite{Brambilla:2004wf,Kogut:1974sn,Wilson:1974sk,Tryon:1972be,Bali:2000gf,Kawanai:2011jt}. The linear confinement potential represents the nonperturbative behavior of QCD forces between the static heavy quarks at long distances and it corresponds to the square-law limit for the Wilson loop. The linear confinement potential is applied not only to the heavy mesons but also to the light mesons \cite{Ebert:2017els,Baldicchi:1998gt,Ebert:2009ub,Godfrey:1985xj,Stanley:1980zm,Fulcher:1994ek,Munz:1993si,
Ricken:2003ua}.

The linear confinement potential is used with different dynamic equations \cite{Godfrey:1985xj,Stanley:1980zm,Eichten:1978tg,Klempt:1995ku,Olsson:1995st,Spence:1993tb}. As discussed in \ref{subsec:dyn}, the Schr\"{o}dinger equation with the linear potential gives the convex {\rts} while the spinless Salpeter equation, the Dirac equation and the Klein-Gordon equation produce the linear {\rts}. The quadratic form of the spinless Salpeter-type equation \cite{Chen:2018hnx,Chen:2018nnr} and the eigenvalue equation for the square mass operator \cite{baldicchi:reg,Baldicchi:2007ic,Baldicchi:2000cf,Brambilla:1995bm,Baldicchi:2007zn} lead to the concave {\rts} which are in good agreement with the experimental data (see Sec. \ref{sec:rts}). Therefore, if the confinement potential is linear, the quadratic form of the spinless Salpeter-type equation and the eigenvalue equation for the square mass operator are preferred according to the concavity of the meson {\rts}.

\subsection{Nonlinear {\rts} from different approaches}
The curvature of the meson {\rts} leads to different inferences for different models. For example, in a string model, the massless string leads to the linear {\rts} while the relativistic string with massive ends can generate the nonlinear {\rts} \cite{Sharov:2013tga,Chen:2017fcs,Afonin:2014nya}.

In Refs. \cite{Sergeenko:1993sn,Sergeenko:1994ck,Ebert:2011jc}, the authors propose an interpolated formula
(\ref{nonlreg}). Without the physical constraints on the parameters, formula (\ref{nonlreg}) can be concave, convex or has the inflection point. As the physical constraints on the parameters are considered, the formulas will be only concave, which cannot describe the convex {\rts}. The interpolated formulas in Eq. (\ref{nrtf}) are also in this case.

In Ref. \cite{Dey:1996xj}, the author obtain the concave {\rts} from the $\kappa$-deformed Poincar\'{e} phenomenology,
\bea\label{reggekap}
M=\frac{2}{\epsilon}\sinh^{-1}\left[\left(\frac{\epsilon}{2}\right)^2\left(\frac{l}{\alpha'}
     +\frac{n_r}{\beta'}+\frac{S}{\gamma'}+\frac{J}{\delta'}\right)
     +\sinh^2\left(\frac{m\epsilon}{2}\right)\right]^{1/2},
\eea
where $\epsilon$ is the inverse of the deformation parameter $\kappa$, $m$ is the quark mass. $M$ is the bound state mass, $J$ is the angular momentum, and $n$ is the radial quantum number. The spin-spin and spin-orbit and tensor effects can be included through the terms involving $J$ and $S$. $\alpha'$, $\beta'$, $\gamma'$ and $\delta'$ are the parameters. Eq. (\ref{reggekap}) is a concave function, which cannot be convex or have the inflection point.

In Ref. \cite{Brisudova:1999ut}, the author discuss the generalized string with the massless ends. They obtain the square-root trajectory
\bea\label{reggfb1}
M^2=\left(\frac{\sigma}{\mu}\right)^2-\left(\frac{\sigma}{\mu}-{\pi\mu}J\right)^2
\eea
from the generalized massless string with the potential $V(\rho)=\sigma/(\pi\mu)\arctan(\pi\mu\rho)$
and the logarithmic trajectory
\bea\label{reggfb2}
M^2=\left(\frac{\sigma}{\mu}\right)^2\left(1-e^{-2\pi{\mu^2}J/\sigma}\right)
\eea
from the generalized massless string with the potential $V(\rho)=\sigma/(2\pi\mu)(2\arctan(2\pi\mu\rho)-\log[1+(2\pi\mu\rho)^2]/(2\pi\mu\rho))$, respectively.
In the above two equations, $\sigma$ is the string tension and $\mu$ is a parameter. Eqs. (\ref{reggfb1}) and (\ref{reggfb2}) are concave functions.

In Ref. \cite{Veseli:1996gy}, one concave formula is given
\bea\label{reggestr2}
M^2=\left(\sqrt{{\pi\sigma}l}+m_Q\right)^2,
\eea
where $m_Q$ is the heavy quark mass. In Ref. \cite{Chen:2017fcs}, the authors give the derivation of Eq. (\ref{reggestr2}) by following the Nambu's picture for hadrons in which quarks are connected by a gluon flux tube. Eq. (\ref{reggestr2}) is concave and cannot be convex or have the inflection point. The formula \cite{Afonin:2014nya}
\bea
M^2=\left(\sqrt{{2\pi\sigma}(n+b)}+2m\right)^2
\eea
is also in this case, where $n$ is the radial quantum number. Other formulas which can give the concave {\rts} are not discussed due to their many parameters \cite{Inopin:1999nf,Sharov:2013tga}.

The curvature of the meson {\rts} can assist in the choice of the appropriate models to describe mesons. Thus it can be taken as a guide in constructing models, because the curvature of the meson {\rts} is model-independent. The appropriate models should give the spectra which can produce the concave {\rts} according to the concavity of the meson {\rts}. If the majority of the meson {\rts} are concave while a few meson {\rts} are convex which have been neither confirmed nor excluded completely, many existing models should be corrected or even be reconstructed, which leads to the further understanding of the meson dynamics.

\section{Complexity of the {\rts}}\label{sec:comp}
The complexity of the {\rts} is included in this section for completeness. The {\rts} are essentially nonlinear complex functions. The linear {\rts} which can happen only for the zero-width resonances will become nonlinear as the width of the resonance pole is taken into account. The real and imaginary parts of the {\rts} are intimately related by analyticity. It is usually assumed that the {\tjs} are real for $t<t_0$. Since $\alpha(t)$ is real analytic a dispersion relation can be written as \cite{Collins:1971ff,Fiore:2000fp,Fiore:2015lnz,Kobylinsky:1976vq}
\bea\label{dispersionr}
\alpha(t)=A(t)+\frac{t}{\pi}\int_{t_0}^{\infty}\frac{{\rm Im}\alpha(t')}{t'(t'-t)}dt'
\eea
if one subtraction is sufficient. We assume ${\rm Re}\alpha(t)\underset{t\to\infty}{\to}A(t)$, where $A(t)=\beta_l^{-3/2}(t-c_l)^{3/2}-b_l$ is obtained from Eq. (\ref{rtint}). ${\rm Re}\alpha(t)$ and ${\rm Im}\alpha(t)$ are the real and imaginary parts of the {\tj} $J=\alpha(t)$ ($J$ is the total spin for mesons), respectively. ${\rm Im}\alpha(t)>0$ for $t>t_0$, where $t_0$ is the $t$-channel threshold.

The {\tj} has the threshold behavior arising from the unitarity \cite{Collins:1971ff,Bugrij:1973ph,Fiore:2000fp,Kobylinsky:1976vq,Barut:1962zz,Gribov:1962pst}
\bea
{\rm Im}\alpha(t)\sim(t-t_0)^{{\rm Re}\alpha(t_0)+1/2}\;{\rm as}\; t{\to}t_0.
\eea
And the {\tj} has the square-root asymptotic behavior on the physical sheet  \cite{Collins:1971ff,Bugrij:1973ph,Fiore:2000fp,Kobylinsky:1976vq,Gribov:1962pst,Degasperis:1970us}
\bea\label{srasymp}
\alpha(t){\sim}(-t)^{1/2}\;{\rm as}\; |t|\to\infty.
\eea
In Refs. \cite{Chen:2018hnx,Chen:2018nnr,Lyubimov:1977km,Brandt:1997gi,Afonin:2014nya,Chen:2017fcs,Veseli:1996gy}, the {\rts} show different behaviors and are used as the effective phenomenological formulas. This kind of cases are similar to the linear {\rt} which is effective and used widely but violates the Froissart bound \cite{Brisudova:1999ut}. However, further investigation is still needed in the future.

\begin{figure*}[!tbp]
\centering
\subfigure[]{\label{fig:subfigure:3a}
\includegraphics[scale=0.8]{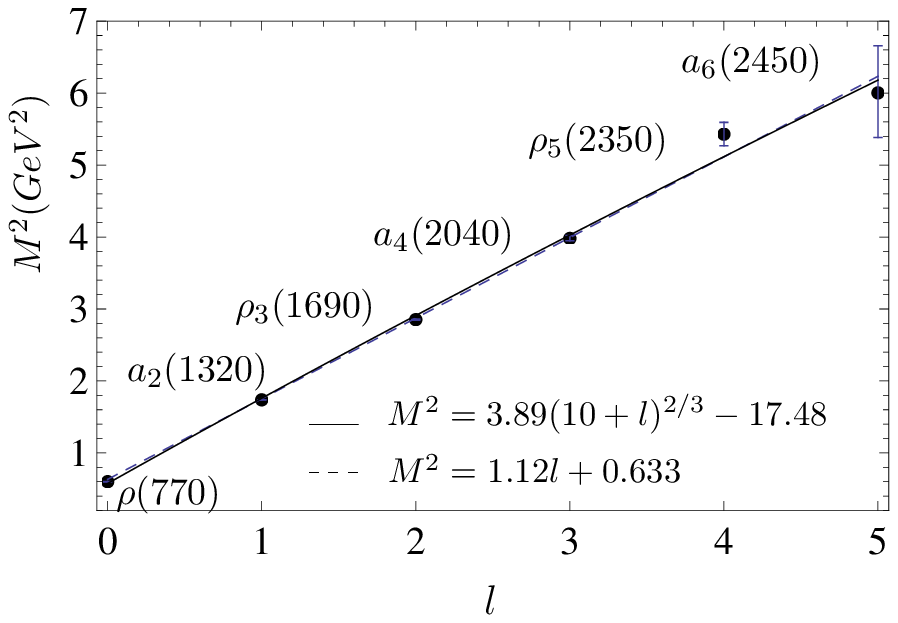}}
\subfigure[]{\label{fig:subfigure:3b}
\includegraphics[scale=0.8]{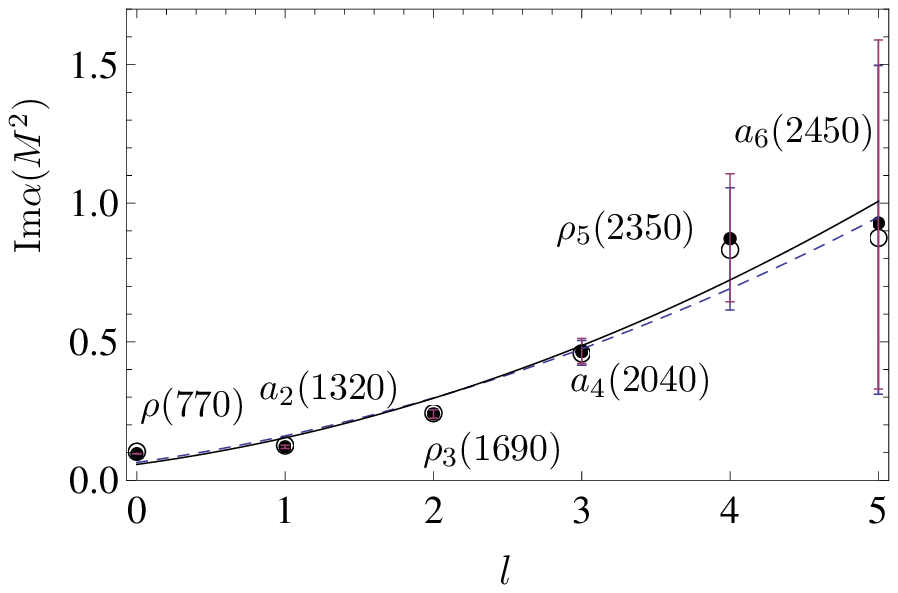}}
\caption{Left panel: The orbital {\rts} for the $\rho/a$ mesons. Right panel: The $(l,{\rm Im}\alpha(M^2))$ plots for the $\rho/a$ mesons. The black and dashed lines are displayed to guide the eye. The circles denote the values obtained by using $M^2=1.12l+0.633$ and Eq. (\ref{widthf}). The solid dotes denote the values yielded by using $M^2=3.89(10+l)^{2/3}-17.48$ and Eq. (\ref{widthf}). $l$ is the orbital angular momentum.}\label{fig:comp}
\end{figure*}

There is a relation between the imaginary part of the {\rt} and the total decay width $\Gamma$ \cite{Collins:1971ff,Fiore:2000fp}
\bea\label{widthf}
\Gamma=\frac{{\rm Im}\alpha(M^2)}{M\alpha'(M^2)},\quad \alpha'(t)=\frac{d{\rm Re}\alpha(t)}{dt},
\eea
where $M$ is the resonance's mass. Since the widths of the states are known one can deduce the corresponding ${\rm Im}\alpha$ from Eq. (\ref{widthf}).
Then the plots of ${\rm Im}\alpha$ against $l$ can be obtained. Some $(l,{\rm Im}\alpha)$ plots are convex, even show the square-like behavior or the square-root-like behavior in the $({\rm Im}\alpha,l)$ plots, see Fig. \ref{fig:comp}. While some $(l,{\rm Im}\alpha)$ plots for mesons are concave.

\section{Conclusions}\label{sec:con}

We have shown by fitting the spectra of a large majority of mesons that the radial and orbital {\rts} for mesons prefer being concave in the $(n_r,M^2)$ and $(l,M^2)$ planes, respectively. The concavity of the meson {\rts} is model-independent. No meson {\rts} having the inflection points are observed. The convex meson {\rts} are neither definitely confirmed nor excluded completely until now. More theoretical analyses and more experimental data are needed to construct the {\rts} which seem convex at present.

The curvature is a fundamental property of the meson {\rts}. The curvature of the meson {\rts} can be used to classify hadrons.
If the convex {\rts} do not exist, the concavity can act as a criterion to choose a newly observed meson or to assign a particle to the unwell-established state.
If the vast majority of the meson {\rts} are concave while a few meson {\rts} are convex which neither have been confirmed nor have been completely excluded at present, mesons should be classified according to different curvature.

The curvature of the meson {\rts} can assist in the choice of the appropriate models to describe mesons. Therefore,
the curvature of the meson {\rts} can be taken as a guide in constructing models. The appropriate models should yield the spectra which can produce the concave {\rts} according to the concavity of the meson {\rts}. If the large majority of the meson {\rts} are concave, whereas a few of the meson {\rts} are convex, many existing models should be corrected or even be reconstructed, which will lead to the further understanding of the meson dynamics.

\section*{Acknowledgements}
We are very grateful to the anonymous referee(s) for the valuable comments and suggestions.

\appendix
\section{{\rts} for the light unflavored mesons}\label{app:light}
\subsection{{\rts} for the light unflavored mesons}
The spectra of the light unflavored mesons are listed in Table \ref{tab:nsmm}. The states are assigned according to Refs. \cite{Ebert:2017els,Masjuan:2012gc,Sonnenschein:2014jwa,Tanabashi2018pdg,Xue:2018jvi,Fischer:2014xha}. We employ Eq. (\ref{rtint}) to fit the spectra of the light unflavored mesons in Table \ref{tab:nsmm}. As shown in Tables \ref{tab:coefr} and \ref{tab:coefo}, there are eight radial {\rts} and ten orbital {\rts} can be described by Eq. (\ref{rtint}) which are concave. There are eight radial {\rts} (for $a_0(980)$, $\rho(1700)$, $\pi_2(1670)$, $\omega(782)$, $f_2(1270)$, $\omega_3(1670)$, $\phi(1020)$ and $\eta'(958)$) and eight orbital {\rts} (for $\pi(1800)$, $\rho(1450)$, $a_1(1260)$, $a_1(1640)$, $\eta$, $\eta(1295)$, $\eta(1760)$ and $\eta'(958)$) look like convex. We notice that these sixteen convex {\rts} are only possibly convex. More theoretical studies and more experimental data are needed to confirm the possible convexity of these {\rts}.

\begin{longtable*}{@{ \ }c@{ \ }c@{ \ \ }c@{ \ }c@{ \ }c@{ \ }c@{ \ }c@{ \ \
    }c@{ \ }c@{ \ }c@{ \ }c@{ \ }}
\caption{Masses of the light ($q=u,d$) unflavored mesons (in MeV).} \label{tab:nsmm}\\
\hline
\hline%\vspace*{-0.5cm}
%\\%[1pt]
%&&Theory
%&\multicolumn{4}{l}{\underline{\hspace{2.1cm}Experiment\hspace{2.1cm}}}&
%Theory&\multicolumn{2}{l}{\underline{\hspace{.6cm}Experiment\hspace{.6cm}}}\\
$n^{2S+1}L_J$    &$J^{PC}$    & $I=1(q\bar{q})$     &PDG\cite{Tanabashi2018pdg}  & $I=0(q\bar{q})$    &PDG\cite{Tanabashi2018pdg}     & $I=0(s\bar{s})$  &PDG\cite{Tanabashi2018pdg}   \\ %[2pt]
\hline
\endfirsthead
\caption[]{(continued)}\\
\hline\hline
%&&Theory
%&\multicolumn{4}{l}{\underline{\hspace{2.1cm}Experiment\hspace{2.1cm}}}&
%Theory&\multicolumn{2}{l}{\underline{\hspace{.6cm}Experiment\hspace{.6cm}}}\\
$n^{2S+1}L_J$    &$J^{PC}$    & $I=1$     &PDG\cite{Tanabashi2018pdg}   & $I=0$    &PDG\cite{Tanabashi2018pdg}    & $I=0$  &PDG\cite{Tanabashi2018pdg}  \\ %[2pt]
\hline
\endhead
\hline
\hline
\endfoot
\endlastfoot
$1^1S_0$& $0^{-+}$ & $\pi^0$      &134.9770$\pm$0.0005      &$\eta$    &       547.862$\pm$0.017 &$\eta'(958)$   &$957.78\pm0.06$\\
$2^1S_0$& $0^{-+}$ & $\pi(1300)$  &1300$\pm$100            &$\eta(1295)$   &1294$\pm$4             &$\eta(1475)$  &1476$\pm$4\\
$3^1S_0$& $0^{-+}$ & $\pi(1800)$  &1812$\pm$12            &$\eta(1760)$   &1751$\pm$15            &$X(1835)$  &$1826.5^{+13.0}_{-3.4}$\\
$4^1S_0$& $0^{-+}$ & $\pi(2070)$  &2070$\pm$35           &$\eta(2010)$   &2010$^{+35}_{-60}$      & $\eta(2225)$ &$2221^{+13}_{-10}$\\
$5^1S_0$& $0^{-+}$ & $\pi(2360)$  &2360$\pm$25            &$\eta(2320)$   &2320$\pm$15            & & \\
\hline %%
$1^3S_1$& $1^{--}$ & $\rho(770)$  &775.26$\pm$0.25        &$\omega(782)$  &782.65$\pm$0.12        &$\phi(1020)$  & 1019.461$\pm$0.016\\
$2^3S_1$& $1^{--}$ & $\rho(1450)$ &1465$\pm$25           &$\omega(1420)$ &1400-1450              &$\phi(1680)$  & 1680$\pm$20\\
$3^3S_1$& $1^{--}$ &  $\rho(1900)$ &1909$\pm$17$\pm$25     &$\omega(1650)$ &1670$\pm$30           &$\phi(2170)$  &2188$\pm$10 \\
$4^3S_1$& $1^{--}$ & $\rho(2150)$  & $2150\pm40\pm50$       &$\omega(1960)$ &1960$\pm$25       &       & \\
$5^3S_1$& $1^{--}$ &               &          &$\omega(2290)$ &$2290\pm20$            &               & \\
\hline %%%
$1^3P_0$& $0^{++}$ & $a_0(980)$   &980$\pm$20          &$f_0(1370)$    &1200-1500              &$f_0(1710)$   &1723$^{+6}_{-5}$ \\
$2^3P_0$& $0^{++}$ &$a_0(1450)$   &1474$\pm$19            &               &        &                    &\\
$3^3P_0$& $0^{++}$ & $a_0(2020)$    &2025$\pm$30  &               &           &               &\\
%$4^3P_0$& $0^{++}$ &&       &               &              &                     &\\
\hline %%%
$1^3P_1$& $1^{++}$ & $a_1(1260)$   &1230$\pm$40         &$f_1(1285)$    &1281.9$\pm$0.5          &$f_1(1420)$  &1426.4$\pm$0.9\\
$2^3P_1$& $1^{++}$ & $a_1(1640)$   &1654$\pm$19        &               &                        &   &\\
$3^3P_1$& $1^{++}$ &$a_1(2095)$    &2096$\pm$17$\pm$121   &               &                       &                    &\\
$4^3P_1$& $1^{++}$ & $a_1(2270)$   &2270$^{+55}_{-40}$    &$f_1(2310)$    &2310$\pm$60            &                     &\\
\hline %%%%
$1^3P_2$& $2^{++}$ & $a_2(1320)$   &1318.3$^{+0.5}_{-0.6}$  &$f_2(1270)$   &1275.5$\pm$0.8         &$f_2'(1525)$ &1525$\pm$5\\
$2^3P_2$& $2^{++}$ & $a_2(1700)$   &1732$\pm$9   &$f_2(1750)$   &1755$\pm$10 & $f_2(1950)$& $1944\pm12$\\
$3^3P_2$& $2^{++}$ &$a_2(1990)$    &2050$\pm$10$\pm$40   &$f_2(2150)$   &$2157\pm12$            &&\\
%$4^3P_2$& $2^{++}$ &                 &            & &         &                      &\\
\hline %%%%
$1^1P_1$& $1^{+-}$ & $b_1(1235)$   &1229.5$\pm$3.2         &$h_1(1170)$   &1170$\pm$20            &$h_1(1380)$ &1407$\pm12$\\
$2^1P_1$& $1^{+-}$ &               &                      & $h_1(1595)$ & $1594^{+18}_{-60}$                      &                    &\\
$3^1P_1$& $1^{+-}$ &$b_1(1960)$    &1960$\pm$35     &$h_1(1965)$   &1965$\pm$45           &                     &\\
$4^1P_1$& $1^{+-}$ &$b_1(2240)$    &2240$\pm$35     &$h_1(2215)$   &2215$\pm$40           &                     &\\
\hline %%%%
$1^3D_1$& $1^{--}$ & $\rho(1700)$  &1720$\pm20$   &$\omega(1650)$ &1670$\pm$30           &                    &\\
$2^3D_1$& $1^{--}$ & $\rho(2000)$ & $2000\pm30$   &               &                    &&\\
$3^3D_1$& $1^{--}$ & $\rho(2270)$  &$2265\pm40$             &               &                    &                    &\\
\hline %%%
$1^3D_2$& $2^{--}$ &               &              &               &                    &                    &\\
$2^3D_2$& $2^{--}$ &$\rho_2(1940)$ &1940$\pm$40  &$\omega_2(1975)$ &1975$\pm$20      &                    &\\
$3^3D_2$& $2^{--}$ &$\rho_2(2225)$ &2225$\pm$35   &$\omega_2(2195)$ &2195$\pm$30      &                     &\\
\hline %%%%%
$1^3D_3$& $3^{--}$ & $\rho_3(1690)$&1688.8$\pm$2.1 &$\omega_3(1670)$ &1667$\pm$4        &$\phi_3(1850)$      &1854$\pm$7 \\
$2^3D_3$& $3^{--}$ &               &               & $\omega_3(1945)$ & $1945\pm20$      &                    &\\
$3^3D_3$& $3^{--}$ &               &      &$\omega_3(2255)$  &2255$\pm$15     &                     & \\
\hline %%%%
$1^1D_2$& $2^{-+}$ & $\pi_2(1670)$ &1672.2$\pm$3.0  &$\eta_2(1645)$ &1617$\pm$5      &$\eta_2(1870)$      &1842$\pm$8\\
$2^1D_2$& $2^{-+}$ & $\pi_2(2005)$ &1974$\pm$14$\pm$83  &$\eta_2(2030)$ &2030$\pm$5$\pm$15     &                    &\\
$3^1D_2$& $2^{-+}$ & $\pi_2(2285)$ &$2285\pm20\pm25$      &$\eta_2(2250)$     &2248$\pm$20           &                    &\\
\hline %%%
$1^3F_2$& $2^{++}$ &               &              &$f_2(1810)$ &1815$\pm$12           &$f_2(2150)$   &2157$\pm$12\\
$2^3F_2$& $2^{++}$ &                &     &$f_2(2140)$ &2141$\pm$12           &                    &\\
$1^3F_3$& $3^{++}$ &$a_3(1875)$    &1874$\pm$43$\pm$96  &               &               && \\
$2^3F_3$& $3^{++}$ &$a_3(2275)$    &2275$\pm$35  &               &                    &                    &\\
$1^3F_4$& $4^{++}$ & $a_4(2040)$   &1995$^{+10}_{-8}$  &$f_4(2050)$    &2018$\pm$11         &                    &\\
$2^3F_4$& $4^{++}$ & $a_4(2255)$   &2237$\pm$5   &&            &                     &\\
$1^1F_3$& $3^{+-}$ & $b_3(2030)$   & $2032\pm12$   & $h_3(2025)$  & $2025\pm20$        &$h_3(2275)$ &2275$\pm$25\\
$2^1F_3$& $3^{+-}$ &$b_3(2245)$    &2245$\pm$50    &               &                    &                     &\\
\hline
$1^3G_3$& $3^{--}$ & $\rho_3(1990)$&1982$\pm$14    &     &      &                    &\\
$2^3G_3$& $3^{--}$ & $\rho_3(2250)$&2260$\pm$20    &     &       &                     &\\
$1^3G_4$& $4^{--}$ &$\rho_4(2230)$&2230$\pm$25   &$\omega_4(2250)$&2250$\pm$30         &                     &\\
%$2^3G_4$& $4^{--}$ &              &               &               &                     &       &\\
%%
$1^3G_5$& $5^{--}$ & $\rho_5(2350)$&2330$\pm$35  &$\omega_5(2250)$&2250$\pm$70       &                    &\\
%$2^3G_5$& $5^{--}$ &              &               &               &                 &        &\\
%%
$1^1G_4$& $4^{-+}$ & $\pi_4(2250)$  &2250$\pm$15  &$\eta_4(2330)$ &2328$\pm$38               &            &\\
%$2^1G_4$& $4^{-+}$ &               &   &         &    &                      &\\
%---------------------------
%\pagebreak
$1^3H_4$& $4^{++}$  &  &    &$f_J(2220)$    & 2231.1$\pm$3.5       &                     &\\
%$1^3H_5$& $5^{++}$  &              &              &               &                    &     &\\
$1^3H_6$& $6^{++}$  &$a_6(2450)$   &2450$\pm$130  &$f_6(2510)$    &2469$\pm$29          &            &\\
%$1^1H_5$& $5^{+-}$  &              &             &               &                    &      & \\ %[2pt]
\hline
\hline
\end{longtable*}

\begin{table}[!htbp]
\caption{Parameters of the radial {\rts} of the form (\ref{rtint}) for the light unflavored mesons.} \label{tab:coefr}
\centering
\begin{tabular*}{0.48\textwidth}{@{\extracolsep{\fill}}cccc@{}}
\hline\hline
                &   $\beta_{n_r}$      &  $b_{n_r}$    & $c_{n_r}$ \\
\hline
 $\pi^0$        &  2.78                &  0.79         &$-$2.36    \\
 $\rho(770)$    &  2.63                &  0.86         &$-$1.77     \\
 $a_1(1260)$      &  3.23                &  3.58         &$-$6.04     \\
 $a_2(1320)$      &  3.52                &  5.96         &$-$9.85     \\
 $b_1(1235)$      &  3.94                &  10.0            &$-$16.79     \\
 $\eta$          &  3.23                &  3.24        &$-$6.78     \\
 $h_1(1170)$     &  4.03                & 10.0           & $-$17.37    \\                                                             $\eta_2(1645)$  &  1.60                & 0.02           & 2.50    \\
\hline\hline
\end{tabular*}
\end{table}

\begin{table}[!htbp]
\caption{Parameters of the orbital {\rts} of the form (\ref{rtint}) for the light unflavored mesons.} \label{tab:coefo}
\centering
\begin{tabular*}{0.48\textwidth}{@{\extracolsep{\fill}}cccc@{}}
\hline\hline
                &   $\beta_{l}$      &  $b_{l}$    & $c_{l}$ \\
\hline
 $\pi^0$        &   2.83             & 1.50        &$-$3.69     \\
 $\pi(1300)$    &   3.76             & 10.0           &$-$ 15.78         \\
 $\rho(770)$    &   3.89             & 10.0       &$-$17.48         \\
 $a_0(980)$    &   1.42             & -1.0       &1.16          \\
  $a_0(1450)$    &   1.41             & -1.0       &2.32          \\
 $\omega(782)$  &   4.16             & 13.0       & $-$22.47     \\
 $\omega(1420)$ &  1.22              & 0.09       & 1.79       \\
 $\omega(1650)$ &  1.49              & 0.0       & 2.89       \\
 $f_0(1370)$    &  2.28              & 4.69        & $-$5.42         \\
  $\phi(1020)$  &  2.36              & 1.37        & $-$1.88         \\
\hline\hline
\end{tabular*}
\end{table}

\subsection{Discussions}
 Twelve {\rts} are constructed by fitting three states and four {\rts} are obtained by fitting more than three states among sixteen possible convex {\rts}. Same as the example discussed in \ref{sub:convex}, twelve {\rts} obtained by fitting three states maybe arise from the insufficient experimental data or the inappropriate assignment of states. The convex radial {\rt} for $\omega(782)$ including five states will become concave if $\omega(2290)$ is excluded which is assigned to the unwell-established $5^3S_1$ state. The radial {\rt} for the $0^{-+}$ $s\bar{s}$ state is convex if $\eta(2225)$ is included \cite{Masjuan:2012gc,Xue:2018jvi} and becomes concave if $\eta(2225)$ is excluded or $X(2500)$ is assigned as the $5^1S_0$ state \cite{Xue:2018jvi,Pan:2016bac}. The orbital {\rt} for the $1^1S_0$ $s\bar{s}$ state will be convex if $\eta'(958)$ is assigned as the $0^{-+}$ $s\bar{s}$ state \cite{Masjuan:2012gc,Xue:2018jvi} while it comes to be concave if $\eta$ is assumed to be the $1^1S_0$ $s\bar{s}$ state \cite{Tanabashi2018pdg}.

The orbital {\rt} for the $0^{-+}$ $q\bar{q}$ state is obtained by fitting the spectra of five states, which is convex whether $\eta$ or $\eta'(958)$ is assigned as the $1^1S_0$ $q\bar{q}$ state \cite{Tanabashi2018pdg,Masjuan:2012gc,Xue:2018jvi}. $h_3(2025)$ and $\eta_4(2330)$ are unconfirmed states among the five states \cite{Sonnenschein:2014jwa}. If $\eta'(958)$ is assigned as the $1^1S_0$ $q\bar{q}$ state \cite{Tanabashi2018pdg}, both the radial and orbital {\rts} for the $1^1S_0$ $q\bar{q}$ state become convex. In summary, the convexity of the orbital {\rt} for the $0^{-+}$ $q\bar{q}$ state is not confirmed.

We can conclude that no convex {\rts} for the light unflavored mesons have been confirmed until now, therefore, the {\rts} for the light unflavored mesons prefer being concave. If some of the {\rts} for meson are convex illustrated by the future analyses, the {\rt} for the $0^{-+}$ $q\bar{q}$ state is likely to be the first one to be confirmed.

\section{Interpolated formulas}

The {\rts} in Eq. (\ref{rtint}) are obtained from the quadratic form of the spinless Salpeter-type equation by employing the Bohr-Sommerfeld quantization approach \cite{brau:04bs,tomonaga}. In the {\rts} in (\ref{rtint}), the long-range potential's contributions are considered while the short-range potential's contributions are neglected. In this section, we consider the contributions from the short-range potential and then present one interpolated form of the {\rts} by incorporating both the long-range contributions and the short-range contributions.

The quadratic form of the spinless Salpeter-type equation reads \cite{Chen:2018hnx,baldicchi:reg,chen:2016gvs,chen:2017qsse,DiSalvo:1994mf,Baldicchi:2007ic,Baldicchi:2000cf,Brambilla:1995bm}
\begin{equation}\label{eq:quadr}
    M^2\Psi({\bf r}) =M_0^2\Psi({\bf r}) +{\mathcal U}\Psi({\bf r}),\;\;M_0=\omega_1+\omega_2,
\end{equation}
where $\omega_i$ is the square-root operator of the relativistic kinetic energy of constituent
\bea
\omega_i=\sqrt{m_i^2+{\bf p}^2}=\sqrt{m_i^2 -\Delta}.
\eea
$m_i$ is the effective mass in the phenomenological model. For simplicity, we assume that ${\mathcal U}$ takes the form
\begin{eqnarray}\label{potsimp}
{\mathcal U} = -{A\over r} + {B}r,
\end{eqnarray}
which is a variant of the well-known Cornell potential \cite{Eichten:1974af}.

In the long range, the linear confinement potential is dominant while the color Coulomb potential is subordinate, then the {\rts} (\ref{rtint}) are obtained. In the shortly distant region, the color Coulomb part will dominate and the linear confinement potential can be ignored. Using Eqs. (\ref{eq:quadr}) and (\ref{potsimp}), we have two auxiliary equations
\begin{align}\label{eq:csch}
    M_1^2\Psi({\bf r}) =& 4({\bf p}^2+m_1^2)\Psi({\bf r}) -\frac{A}{r}\Psi({\bf r}),\nonumber\\
    M_2^2\Psi({\bf r}) =& 4({\bf p}^2+m_2^2)\Psi({\bf r}) -\frac{A}{r}\Psi({\bf r}).
\end{align}
The obtained eigenvalues read, respectively,
\begin{align}\label{eigencom}
M_1^2=& 4m_1^2-\frac{A^2}{16}\frac{1}{(n_r+l+1)^2},\nonumber\\
M_2^2=& 4m_2^2-\frac{A^2}{16}\frac{1}{(n_r+l+1)^2}.
\end{align}
Because $4\omega_1^2\le(\omega_1+\omega_2 )^2\le4\omega_2^2$ if $m_1{\le}m_2$,
the eigenvalues of Eq. (\ref{eq:quadr}) in the short-range case are expected to be of the form
\bea\label{eigencoloum}
M^2= 4{\tilde{m}}^2-\frac{{\tilde{A}}^2}{(n_r+l+1)^2},
\eea
where $\tilde{m}$ and $\tilde{A}$ are parameters. Combining Eqs. (\ref{rtint}) and (\ref{eigencoloum}), we propose the parameterized formulas
\begin{align}\label{nrtf}
M^2=&\frac{\gamma_1}{(l+n_r+1)^2}+\beta_1(l+b_1)^{2/3}+c_1, \nonumber\\
M^2=&\frac{\gamma_2}{(l+n_r+1)^2}+\beta_2(n_r+b_2)^{2/3}+c_2.
\end{align}
The first terms are from the color Coulomb potential and the second terms are from the linear confinement potential. Therefore, $\gamma_1$ and $\gamma_2$ should be negative while $\beta_1$ and $\beta_2$ should be positive. Different from the {\rts} (\ref{rtint}) which are concave, the formulas in (\ref{nrtf}) can be concave or convex, and can have the inflection points as the physics constraints on the parameters are not considered. The physical constraints on the parameters make the formulas in (\ref{nrtf}) be concave.

%-========================================================
%\section*{References}

\end{document}